\documentclass[10pt,conference]{IEEEtran}
\usepackage{graphicx} %
\usepackage{amsmath}
\usepackage{amssymb}
\usepackage{url}
\usepackage{listings}

\usepackage{booktabs} 
\usepackage{tabularx}
\usepackage[para,online,flushleft]{threeparttable}

\usepackage{hyperref}

\newcommand{\schemename}{\textsc{AutoStub}}

\usepackage{tcolorbox}
\tcbuselibrary{skins}
\newtcolorbox{answerbox}{
  colback=black!5!white,
  colframe=black!70!white,
  fonttitle=\bfseries,
  sharp corners,
  rounded corners,
  boxrule=0.3mm
}

\title{
\schemename: Genetic Programming-Based Stub Creation for Symbolic Execution
}
\author{
    \IEEEauthorblockN{
        Felix Mächtle, 
        Nils Loose, 
        Jan-Niclas Serr, 
        Jonas Sander, 
        Thomas Eisenbarth
    }
    \IEEEauthorblockA{
        Institute for IT Security, University of Luebeck, Germany \\
        Email: \{f.maechtle, n.loose, j.serr,  j.sander, thomas.eisenbarth\}@uni-luebeck.de
    }
}
\date{}

\begin{document}

\maketitle

\begin{abstract}
Symbolic execution is a powerful technique for software testing, but suffers from limitations when encountering external functions, such as native methods or third-party libraries. 
Existing solutions often require additional context, expensive SMT solvers, or manual intervention to approximate these functions through symbolic stubs.
In this work, we propose a novel approach to automatically generate symbolic stubs for external functions during symbolic execution that leverages Genetic Programming. 
When the symbolic executor encounters an external function, \schemename{} generates training data by executing the function on randomly generated inputs and collecting the outputs. Genetic Programming then derives expressions that approximate the behavior of the function, serving as symbolic stubs. These automatically generated stubs allow the symbolic executor to continue the analysis without manual intervention, enabling the exploration of program paths that were previously intractable.
We demonstrate that \schemename{} can automatically approximate external functions with over 90\% accuracy for 55\% of the functions evaluated, and can infer language-specific behaviors that reveal edge cases crucial for software testing.
\end{abstract}

\section{Introduction}

Symbolic execution is a foundational technique in software testing and analysis, enabling exhaustive exploration of program paths. %
Instead of executing the program with specific inputs, symbolic execution represents inputs with symbols and computes expressions over these symbols, generating symbolic expressions that capture the behavior of the program along different execution paths~\cite{DBLP:journals/cacm/King76}. This results in logical expressions representing the constraints under which each path is feasible.
By leveraging constraint solvers to systematically explore  execution paths, symbolic execution facilitates automatic test case generation, bug finding, and formal verification. However, a significant limitation of symbolic execution arises when the analyzed program invokes \emph{external functions}~\cite{DBLP:journals/csur/BaldoniCDDF18/external_functions}, such as native methods, third-party library calls, or uninstrumented functions. These act as black boxes, hindering the symbolic execution by introducing unknown behaviors that cannot be symbolically evaluated. Therefore, any statements that depend on the outputs of such functions become unanalyzable.

For illustration, consider the example in Listing~\ref{fig:intro:example-program}. To test any path dependent on the \texttt{if} statement, symbolic execution  generates expressions representing the relationship between the input variable \texttt{user\_input} and the boolean result of \texttt{verify\_input}. %
However, if \texttt{verify\_input} represents an external function, its internal behavior is inaccessible to the symbolic execution and no relationships can be created.
Consequently, any code that relies on the result of this function, such as the subsequent call to \texttt{display("Access Granted")}, cannot be accurately analyzed unless additional handling or modeling of the external function is provided.

\renewcommand{\figurename}{Listing}
\begin{figure}
\begin{lstlisting}[basicstyle=\footnotesize]
function main(user_input):
    if verify_input(user_input) == True:
        display("Access Granted")
\end{lstlisting}  
\caption{Illustrative example of symbolic execution encountering an external function. Without knowledge of \texttt{verify\_input}, the analysis is unable to fully explore all program paths.}
\label{fig:intro:example-program}
\end{figure}
\renewcommand{\figurename}{Figure}
\setcounter{figure}{0} %

Current approaches to address external functions either rely on additional context information~\cite{DBLP:conf/icse/ZhaiHMZTZQ16/JavaDoc}, expensive SMT solvers~\cite{DBLP:conf/sigsoft/MechtaevGCR18, DBLP:conf/wcre/QiSQZZR12} or manual intervention~\cite{DBLP:conf/tacas/LooseMSE24}, where developers manually provide \emph{symbolic stubs} to approximate the behavior of external functions. %
We explore a different path and introduce \schemename, a novel approach that automates the generation of symbolic stubs for external functions during symbolic execution. We leverage Genetic Programming, a type of machine learning that discovers expressions best fitting inputs to outputs from observed data, making it suitable for approximating the functionality of black-box methods.
When the symbolic executor encounters an external function, \schemename{} generates training data by executing the external function on randomly generated inputs and collecting the corresponding outputs. Genetic Programming is then used to derive expressions that reproduce the relationship between inputs and outputs. These expressions serve as symbolic stub for the external function. This workflow can be seamlessly integrated into the symbolic execution process, allowing the executor to continue the analysis without manual intervention. Hence, \schemename{} allows exploring program paths that were previously untractable due to external functions. To the best of our knowledge, \schemename{} is the first system capable of generating symbolic stubs using Genetic Programming.
All implementation details and datasets are available in our GitHub repository: \url{https://github.com/UzL-ITS/AutoStub}.

\noindent To summarize our contributions are:
\begin{itemize} 
    \item We demonstrate that external functionalities can be approximated using Genetic Programming, enabling automated generation of symbolic stubs across multiple primitive data types and strings.
    
    \item We create a benchmark dataset consisting of 2730 small programs to evaluate correct external functionality handling in symbolic execution.
    
\end{itemize}

\section{Genetic Programming}
Genetic Programming (GP) is an evolutionary computation technique that uses the principles of natural selection and genetics to automatically evolve computer programs to solve complex problems. 
Candidate programs are typically represented as hierarchical tree structures encoding expressions or code segments. These programs evolve over successive generations through genetic operators such as selection, crossover, and mutation, guided by a fitness function that measures their performance on a given task.

The objective is to find a program $P(x)$ that  accurately maps inputs $x$ to an output variable $y$, based solely on observed data. Our approach explores a vast space of expressions constructed from a predefined set of operators, constants, and input variables.
For example, given a training dataset consisting of input-output pairs $(x_y, y_i)$ where $y_i = x_i^2$, one could use GP to find the program $P(x) = x^2$ by searching for combinations of operations and variables that minimize the error between the predicted and actual outputs.

Grammar-Guided Genetic Programming (G3P)~\cite{whigham1995grammatically} is an extension that integrates formal grammars into the GP framework. By defining a context-free grammar that specifies the syntax and allowable constructs, G3P ensures that all generated programs are syntactically correct. This method is particularly useful when evolving programs that involve multiple data types or need to adhere to specific syntactic rules, as it guides the genetic operators to produce valid offspring while exploring the search space more efficiently.

\section{\schemename}
Grammar-Guided Genetic Programming (G3P)~\cite{whigham1995grammatically} is used in \schemename{} to generate symbolic stubs from input-output data.
This method allows the system to seamlessly handle multiple data types while maintaining type consistency throughout the generation process.
A comprehensive set of operators serves as building blocks for symbolic expressions. Specifically, a subset of 40 operators from the SMT-Lib Standard~\cite{barrett2010smt} is used, selected to cover the majority of Java scenarios. These operators include mathematical, logical, and string manipulation functions.

To ensure type consistency across generated expressions, we defined a grammar and used it in conjunction with G3P to generate expression trees of operators.
The fitness function used for evaluation is tailored to the output type of each function: for numeric data types, the Normalized Root Mean Squared Error (NRMSE) measures the approximation accuracy; for Boolean outputs, the classification accuracy is used; and for string outputs, the Levenshtein distance is used. Selection is performed using tournament selection, where multiple candidates compete and only the best one is chosen.
Mutation is accomplished by replacing certain parts of the expression tree with newly generated sub-expressions, while crossover is implemented via one-point crossover, where a randomly selected subtree from one parent is exchanged with a corresponding subtree from the other parent. This allows the offspring to inherit characteristics from both parents.

\subsection{Input Generation}
\label{sec:input-generation}

To generate diverse and representative inputs, we employ a stratified sampling approach tailored for different data types. For integer types (\texttt{byte}, \texttt{short}, \texttt{int}, \texttt{long}), we first randomly select a bit-length $n$ within the allowed range of the data type, effectively dividing the entire range into intervals. %
For each selected bit length, we uniformly generate a random number between $0$ and $2^n - 1$. For signed integers, we randomly assign a sign. %
This method ensures coverage across different magnitudes, ensuring that both small and large values are represented, which is crucial for effectively approximating functions with varying behaviors across different scales.
For floating-point types (\texttt{float}, \texttt{double}), a similar stratified sampling technique is used. First, a random exponent is selected within the allowed range using the technique described for numeric values.
Then a random mantissa with a random sign is assigned.
For strings, random sequences of characters of variable length are constructed. Boolean values are randomly assigned. In addition, special values are included with a 5\% probability, such as \textit{NaN}, \textit{Infinity}, the \textit{max/min} value of the data type, and 0, 1, or -1. Generating 10000 samples took an average of 13ms per method.

\subsection{Datasets}

\noindent\textbf{Expression Dataset.} 
To train and evaluate the correctness of the generated stubs, we utilized all functions from Java internal libraries~\cite{oracleJavalangJava}, specifically focusing on classes related to primitives and mathematical operations. Support for these functions is essential for successful symbolic execution, making them an ideal target. We use all functions from \textit{java.util.*}, where * represents a primitive or a string type, as well as functions from \textit{java.lang.Math} and \textit{java.lang.StrictMath}. We extracted all methods from these classes and filtered them based on specific criteria: the methods must return a primitive or a string, their parameters must be primitives or strings, they must not have side effects on the caller, and they must not throw errors during execution in our input generation pipeline.
From a total of 654 methods in the packages, 273 met the specified criteria. For each of these selected methods, random input values were generated to record their corresponding outputs.
\label{sec:method:scope}

\noindent\textbf{Symbolic Execution Dataset.} 
\label{sec:method:java_benchmark_dataset}
To evaluate the effectiveness of \schemename{} in the context of symbolic execution, we constructed an additional benchmark dataset encompassing test cases for all functions in the aforementioned scope. Similar to Listing \ref{fig:intro:example-program}, the goal of each test case is to verify whether an input passed to an external function results in the desired output. 
For every external function $f$, we generated two random input values and observed the corresponding output values ($o_1, o_2$). We repeated this process until we found two distinct outputs ($o_1 \neq o_2$). Using these outputs, we created a benchmark dataset consisting of Java classes that check whether the external function $f$ returns the desired output values ($o_1, o_2$) for given inputs.

Each Java class reads its arguments ($x_1, x_2$) and invokes the external function to obtain $y_1 = f(x_1)$ and $y_2 = f(x_2)$.
These outputs are then compared against the predetermined values ($y_1 = o_1 \land y_2 = o_2$). As both values ($o_1, o_2$) are distinct, this setup ensures that the test cannot be passed using trivial solutions like a fixed value. 
We adopted this conservative approach to avoid inflated success rates; for example, the function \texttt{isNaN(double): boolean} returns \texttt{false} for all but one input, so testing only one output could be misleading.
To enhance test granularity and increase the number of scenarios, we create ten such tests for each function.

\section{Experiments}

In order to evaluate \schemename{} we propose the following Research Questions:
\begin{itemize}

\item RQ1: How accurately does \schemename{} approximate a function's input-output behavior? %

\item RQ2: How effective are the generated expressions for symbolic execution? %

\end{itemize}

\begin{figure}[t!]
    \centering
    \includegraphics[width=\linewidth]{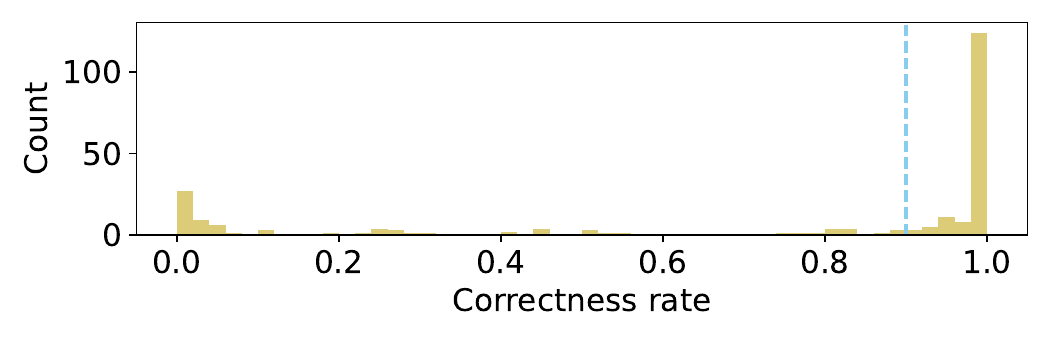}
    \caption{Accuracy distribution of generated expressions over $10^6$ runs, with 55\% of expressions achieving over 90\% correctness in predictions.}
    \label{fig:eval:relative_errors_of_all_expressions}
\end{figure}

\subsection{Accuracy of Generated Expressions}
To address \textbf{RQ1}, we evaluated the accuracy of the generated expressions. Using our input generation strategy, we created $10^6$ new input-output pairs for evaluation. Each generated expression was executed on these inputs, and we recorded the number of times the output matched the expected value. The results, shown in Figure~\ref{fig:eval:relative_errors_of_all_expressions}, show that while some methods rarely returned the correct value, a significant portion performed well. Specifically, 55\% of the expressions returned the correct value more than 90\% of the time.
To establish a lower bound on performance, we also evaluated a random baseline where fitness values were randomly assigned, resulting in 15\% of the generated formulas achieving over 90\% accuracy, demonstrating that even naive approaches can stumble upon partial solutions, but the targeted search strategy in our system significantly outperforms random chance.

Upon manual inspection of the expressions, we found that 79 methods were fully correct. Out of these, 28 instances simply returned the identity value, such as \texttt{Boolean.booleanValue(boolean): boolean}. In 8 instances, the expressions returned a value closely related to the identity, e.g., \texttt{Integer.doubleValue(integer): double}. 
Additionally, 28 instances corresponded directly to a single operator, like an integer multiplication. Finally, 14 instances captured more advanced functionalities, such as checking whether a value is \texttt{NaN}. %
These categories show that while most of the approximated functions are relatively straightforward, some achieve more advanced %
approximations.

To further demonstrate its efficacy, we highlight one function where \schemename{} produced a particularly insightful approximation, \texttt{Double.isNaN(double)}:
This method checks whether a given double value is \texttt{NaN} (Not-a-Number). The generated expression is $!(-1 < | x |)$.
It effectively captures the behavior that any comparison operation involving \texttt{NaN} returns \texttt{false} in Java. By using the absolute value and a comparison to \texttt{-1.0}, the expression exploits the unique properties of \texttt{NaN} to determine its presence without explicit knowledge of the \texttt{Double.isNaN} method. 
This approximation is particularly impressive because it infers language-specific behavior through Genetic Programming. Such insights are particularly valuable in software testing, as they can reveal edge cases and unintended behavior in software systems.

\begin{answerbox}
\schemename{} can automatically approximate external functions, achieving over 90\% accuracy for 55\% of the functions evaluated. %
\end{answerbox}

\begin{figure}
    \centering
    \includegraphics[width=1\linewidth]{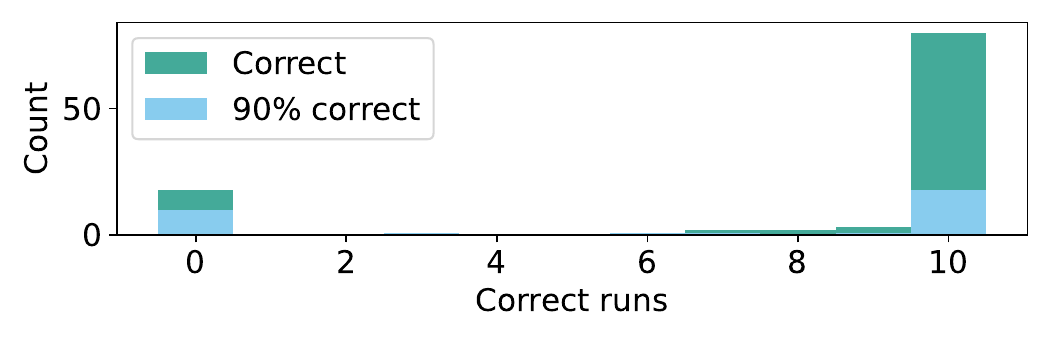}
    \caption{Number of benchmark runs successfully solved during symbolic execution. Zero indicates no solves for this expression whereas ten indicates that the generated expression successfully solved all test cases.}
    \label{fig:eval:symbolic-exec-success}
\end{figure}

\subsection{Symbolic Execution with Generated Expressions}
To answer \textbf{RQ2}, which investigates whether the generated expressions can be employed in symbolic execution, we utilized the benchmark dataset described in  Section~\ref{sec:method:java_benchmark_dataset}. We selected all expressions from RQ1 that achieved an accuracy higher than 90\% and integrated them into SWAT~\cite{DBLP:conf/tacas/LooseMSE24}, a symbolic execution engine for Java. Using SWAT, we symbolically executed all test cases in the benchmark dataset, employing the generated expressions as symbolic stubs for the external functions.
As illustrated in Figure~\ref{fig:eval:symbolic-exec-success}, the majority of samples were fully solvable using the generated expressions with a timeout of one second. On average, each expression required 0.04 seconds to solve, with the maximum solving time being 0.41 seconds on an AMD Ryzen 7 7735U.

A small subset of nine of the generated functional expressions occasionally caused benchmarks to fail, either due to timeouts or because the expressions missed certain edge cases. For example, our enforced timeout proved insufficient for some computations, such as the multiplication of two large integers ($x \times y = 391{,}768{,}351{,}037{,}400{,}960$). In other cases, the solutions found were locally helpful but not entirely accurate, leading to failures.
A third group of functions could not be solved correctly. These functions primarily involved operations such as checking for \textit{NaN} or \textit{Infinity}. The issue arises because Java and the SMT solver used, namely Z3, handle those numbers differently. While \schemename{} successfully identified these relations within the Java context (see RQ1), Z3 handles these specific semantics differently. This discrepancy means that Z3 cannot accurately interpret the language-specific behaviors associated with these values. Therefore, although symbolic stubs can be directly utilized for symbolic execution, expressions that depend on language-specific semantics must be approached with additional care and may require specialized handling within the solver.

\begin{answerbox}
The generated symbolic stubs allow the exploration of program paths that were previously intractable due to external functions, as long as the SMT solver is able to accurately handle the language-specific semantics involved.
\end{answerbox}

\section{Limitations}

\schemename{} is currently limited to stateless functions. %
For example, classes like \texttt{StringBuilder} which retain an internal state, fall outside the scope. This limitation arises because \schemename{} is designed to approximate functions based solely on their input-output behavior in isolation. %
Extending \schemename{} to handle stateful objects remains an open challenge for future work. A possible approach is to build test cases that eventually reduce stateful objects to primitives (e.g., the final output of \texttt{StringBuilder.toString()}). Instead of approximating a single function, a sequence of calls would be approximated to the final output, which can then be compared to the expected result. The internal state must then be passed between calls, e.g., as an argument. %

Additionally, the generated symbolic stubs are limited to expressions of regular complexity within the Chomsky hierarchy. In contrast, the functions we approximate may exhibit Turing-complete behavior, such as loops or recursion. 
This restriction is an intentional design choice to ensure that the generated expressions remain computationally simple, thereby allowing for fast solving. %

\section{Related Work}
Handling external functions in symbolic execution is a well-recognized challenge, and various methods have been proposed to address this issue.
One common method is the use of manual stubs to approximate the behavior of external functions. Developers create symbolic stubs for external functions, allowing the symbolic executor to proceed with the analysis~\cite{DBLP:conf/tacas/LooseMSE24}. While effective, this approach is time-consuming and error-prone, as it requires significant manual effort and may lead to incomplete analysis if some functions are overlooked.

To automate the generation of stubs, SMT-based solutions have been proposed~\cite{DBLP:conf/sigsoft/MechtaevGCR18, DBLP:conf/wcre/QiSQZZR12}. In these approaches, constraints that describe the external functionality are generated, and an SMT solver is used to obtain a model of the external function. However, these methods rely on expensive SMT solving, which can be computationally intensive and may not scale well for complex functions.
As a cheaper alternative, rule-based techniques have been explored~\cite{DBLP:conf/icse/JeonQFFS16/related-work-pasket, DBLP:conf/soict/NguyenO22}. While this approach is computationally cheaper, it is limited by the predefined rules, which may cover only a narrow portion of the possible behaviors of external functions. 

In Java, Zhai \emph{et al.}~\cite{DBLP:conf/icse/ZhaiHMZTZQ16/JavaDoc} used natural language processing (NLP) techniques to match Javadoc comments with code templates, to generate symbolic stubs for internal data structures like lists and sets.
However, this approach requires additional context in the form of Javadocs and focuses primarily on data structures, whereas our work concentrates on methods of internal data types and does not need extra information.

\section{Conclusion}

This work presents \schemename{},  a novel approach that leverages Genetic Programming to automatically generate symbolic stubs for external functions encountered during symbolic execution. Our method addresses a limitation in symbolic execution by enabling the analysis of program paths that involve external functions without requiring manual intervention or extensive contextual information. Experiments show that \schemename{} effectively approximates external functions, achieving over 90\% accuracy for 55\% of evaluated functions and integrating seamlessly with symbolic execution engines. 

Future work should improve the accuracy of the approximations and  extend \schemename{} to support stateful objects. %

\section*{Acknowledgements}
Generative AI was utilized during programming, editing, and grammar enhancement of this work. This work has been supported by funding from the \textit{Agentur für Innovation in der Cybersicherheit GmbH} (Cyberagentur, project \href{https://sovereign-project.de/}{\textit{SOVEREIGN}}) and \textit{Federal Ministry of Research, Technology and Space} (BMFTR, project \href{https://samsmart.de/en/}{\textit{SAM Smart}}).

\bibliographystyle{plain} %
\bibliography{Bib}

\end{document}